# Probing the Interatomic Potential of Solids by Strong-Field Nonlinear Phononics

A. von Hoegen[1], R. Mankowsky[1], M. Fechner[1], M. Först[1], A. Cavalleri[1,2]

[1]Max Planck Institute for the Structure and Dynamics of Matter, 22761 Hamburg, Germany.
[2]Department of Physics, University of Oxford, Clarendon Laboratory, Oxford OX1 3PU, UK.

**Femtosecond optical pulses at mid-infrared frequencies[1] have opened up the nonlinear control of lattice vibrations in solids[2]. So far, all applications have relied on second order phonon nonlinearities[3], which are dominant at field strengths near 1 MVcm$^{-1}$. In this regime, nonlinear phononics can transiently change the average lattice structure, and with it the functionality of a material[4-10]. Here, we achieve an order-of-magnitude increase in field strength, and explore higher-order lattice nonlinearities. We drive up to five phonon harmonics of the $A_1$ mode in LiNbO$_3$. Phase-sensitive measurements of atomic trajectories in this regime are used to experimentally reconstruct the interatomic potential and to benchmark ab-initio calculations for this material. Tomography of the Free Energy surface by high-order nonlinear phononics will impact many aspects of materials research, including the study of classical and quantum phase transitions.**

In the experiments reported here, the highest frequency $A_1$ mode of LiNbO$_3$ was excited with mid-infrared femtosecond pulses, with peak field strengths up to 20 MV/cm. These pulses were tuned to 17.5 THz, slightly to the red of the TO phonon frequency ($\omega_{TO}$ = 19 THz)[11,12]. A time-delayed, 30-fs-long probe pulse at 800-nm wavelength was used to sample these lattice dynamics. We recorded both the time-dependent probe polarization rotation (PR) and second harmonic (SH) intensity (Fig.1c and d). The polarization rotation measured time-dependent changes of the dielectric permittivity of the crystal $\varepsilon_r(\tau)$, whereas the second harmonic sampled the changes in the optical second-order susceptibility $\chi^{(2)}(\tau) \propto \frac{\partial \chi^{(2)}}{\partial Q}$ [13,14], and with it the polar component of the lattice motion. The stable absolute carrier-envelope-phase[1] of the pump field made it possible to directly follow the atomic trajectories in a pump-probe geometry (see Fig. 1b). Spectral interferometry between the frequency upshifted probe and a local oscillator derived from the same probe pulse was used to detect both phase and amplitude of these dynamics[15,16]. Details of the experimental setup and detection process are found in the methods section and Extended Data Fig. 1-3.

As shown in Fig. 1a, the $A_1$ mode involves rotations of the oxygen octahedra, accompanied by c-axis motions against the Nb and Li sublattices (see Fig. 1a). Due to the broken inversion symmetry of the crystal, the $A_1$ mode is both Raman and infrared active[11,12], with a net electric dipole moment along the c-axis.

For small amplitude excitation (0.1 MV/cm), both PR and SH measurements yielded harmonic oscillations (see Fig. 1c and d dashed lines), which were readily attributed to a combination of a 15-THz phonon-polariton and the 19-THz TO phonon of the $A_1$ mode[17]. As shown in Fig. 2, the pump-probe spectrum of the small-field response is well understood by considering the phase-matching between the probe light and the phonon-polariton propagating into the crystal[18, 19]. The dispersion of the phonon-polaritons in LiNbO$_3$ (black line) is plotted as $\omega_p = \frac{c_0}{\sqrt{\varepsilon(\omega)}} q$, where $c_0$ is the speed of light in vacuum and $\varepsilon(\omega)$ the dielectric function. The two light lines $\omega = v_g q$ of the 800 nm (red) and 400 nm (blue) probe ($n_{g,800}$ =2.30, $n_{g,400}$ =3.03)[20] are also shown, where $v_g$ and $q$ denote the group velocity and wave number of the probe light, respectively. Phase-matching occurs at those frequencies for which the light lines intersect the phonon-polariton dispersion curve[18,21], i.e. at 15 THz (PR) and 16 THz (SH).

At high pump fields (20 MV/cm, Fig. 1c and d solid lines), a strongly anharmonic response was observed, with asymmetric oscillations in both PR and SH signals. Figure 3a and b display the corresponding amplitude spectra. In addition to the 15 and 19 THz frequency components observed in the linear response, harmonics of these two frequencies appeared in both PR and SH spectra. The most pronounced peaks were found at multiples of the 15-THz phonon-polariton mode, clearly visible up to n = 5 (75 THz). Correspondingly, the amplitudes of the first three harmonics at $\omega$ = 15, 30 and 45 THz displayed a linear, quadratic and cubic dependence on the excitation field (s. Fig.2 c). The PR spectrum also exhibited peaks at the sum and difference frequencies of these harmonics, likely descending from nonlinear mixing of the TO phonon and the polariton modes (see Extended Data Fig. 4 for a detailed assignments of all peaks). Note that these results are reminiscent of what has been extensively reported in the literature in the context of non-resonant THz and mid-IR harmonic generation[22-27]. However, in the present case the harmonics appear at multiples of the phonon-polariton, instead of the central frequency of the optical pump field, indicating a fundamentally different physical origin.

To analyze these data, we first consider the local lattice response. We start from the anharmonic lattice potential of the driven mode at $\omega_{TO}$, and ignore phonon-polariton propagation. Ab initio density functional theory (DFT) calculations yielded an expression for the anharmonic lattice potential (see methods section)

$$V(Q_{IR}) = \frac{1}{2}\omega_{TO}^2 Q_{IR}^2 + \frac{1}{3}a_3 Q_{IR}^3 + \frac{1}{4}a_4 Q_{IR}^4, \qquad (1)$$

where $a_3$ and $a_4$ are the coefficients of the cubic and quartic potential term (see top panel of Fig. 4). The equation of motion for this mode, when driven by a light pulse with carrier frequency $\omega_{TO}$ and duration $T$, is then given by

$$\ddot{Q}_{IR} + 2\gamma \dot{Q}_{IR} + \omega_{TO}^2 Q_{IR} + a_3 Q_{IR}^2 + a_4 Q_{IR}^3 = Z^* E(t) \qquad (2)$$

where $Z^*$ denotes the effective charge of the phonon mode, $\gamma$ is a dissipation term, and $E(t) = E_0 \sin(\omega_{TO} t) \cdot \exp(-4\ln 2 \cdot t^2 / T^2)$ the excitation pulse profile. The calculated dynamics at field strengths comparable to our experiment, reported in Fig. 4, shows peaks at harmonics of the fundamental frequency $\omega_{TO}$. As expected[28,29], the strongly driven overtones are also slightly red-shifted, an effect that increases with harmonic order (see Extended Data Fig. 5).

A more comprehensive description of our experimental observations was obtained with finite difference time-domain simulations of phonon-polariton propagation[30], which are reported in Fig. 5. In these simulations, we combined the linear optical properties of LiNbO$_3$ with the nonlinear lattice potential of Eq. (1) (see methods and Extended Data Fig. 6 and Table 1). Figure 5a displays the amplitude of the propagating electric field as a function of sample depth and time. Both the phonon-polaritons and the broadband radiation emitted from the anharmonic polar motions propagate from the surface into the bulk following the dispersion imposed by the material. By integrating the simulated electric field along the 800 nm light line, we extracted the response shown in Fig. 5b, yielding good agreement with the polarization-rotation measurement. Figure 5c displays the corresponding amplitude spectrum obtained by Fourier transformation, which comprises peaks at all sum and difference frequencies of the polariton and the TO mode, also in good agreement with the experiment (cf. Fig. 2a).

We next turn to the key results of this paper, which are extracted from the time-dependent changes in the second harmonic intensity $I_{SH}(\tau)$. As discussed in references 13 and 14, changes in $I_{SH}$ induced by a coherent phonon of frequency $\Omega$, can be described as a hyper-Raman scattering process, which involves the generation of sidebands with frequency offset $\Omega$. The signal on the detector involves a frequency integral over the hyper-Raman sidebands at all phonon harmonics, and is therefore linearly proportional to $\Omega \cdot Q(\tau)$ [31,32]. For a sinusoidal field $Q(\tau)$, $\Omega \cdot Q(\tau)$ is proportional to the time derivative $\dot{Q}(\tau)$. To compare simulations with experiments, we integrate the time derivative of the calculated phonon field $Q$ along the blue line of Fig. 5d over the first 2 $\mu m$ beneath the surface, where the SH light is being generated[10] (see Fig. 5e and f). We find good agreement between measured and simulated SH response, which both display broad peaks at multiples of 16 and 19 THz.

Crucially, from the knowledge of $Q(\tau)$ and $\dot{Q}(\tau)$, we can reconstruct the microscopic lattice potential explored during each oscillation cycle. Assuming that the envelope of the driving pulse varies slowly within one phonon period $T$, and that the damping rate $\gamma \ll 1/T$, we can approximate the total energy of the lattice to be a constant $E$ during each cycle, $E = U(Q) + E_{kin}(Q) = const.$ Because $E_{kin}(\tau) = \frac{1}{2}\dot{Q}(\tau)^2$ is known to a proportionality factor, we can retrieve the instantaneous potential energy at each time delay $U(\tau) = E - E_{kin}(\tau)$. By integrating the experimental data over time we recover the corresponding phonon coordinate $Q(\tau)$ and obtain the time independent lattice potential energy $U(Q)$. Furthermore, different cycles with different amplitudes and different total energy $E$ trace fractions of the potential energy $U(Q)$ many times, making its reconstruction highly over-determined.

Figure 6 compares the experimentally reconstructed energy potential of the highest frequency $A_1$-mode of LiNbO$_3$ (filled circles) with the potential obtained from DFT calculations (grey line). The unknown proportionality factor mentioned in the derivation above was chosen from a fit of the experimental data to this calculated potential. More details of this procedure are found in the methods section. We find excellent agreement between the potentials up to the highest excursions in the strongly anharmonic regime reached in our experiment. Hence, strong-field nonlinear phononics can be used to reconstruct interatomic potentials of solids, yielding information not easily retrieved with any other technique.

The tomography of the force field discussed above is straightforwardly extensible to all materials with a large bandgap, like ferroelectrics, for which acceleration of quasi-particles in the field is neglected to first order. A full reconstruction of the force field of a material with N atoms requires the measurement of 3N-3 lattice modes without symmetry considerations. Recent advances in the generation of mid infrared and THz pulses that are both widely tunable and intense[33], make these prospects realistic. Tomographic measurements of force potentials in the vicinity of equilibrium phase transitions will yield crucial information not accessible otherwise. Finally, as the sampling of the potential can be retrieved within one cycle of the pump light, one could envisage measurements of rapidly evolving potential energy surfaces.

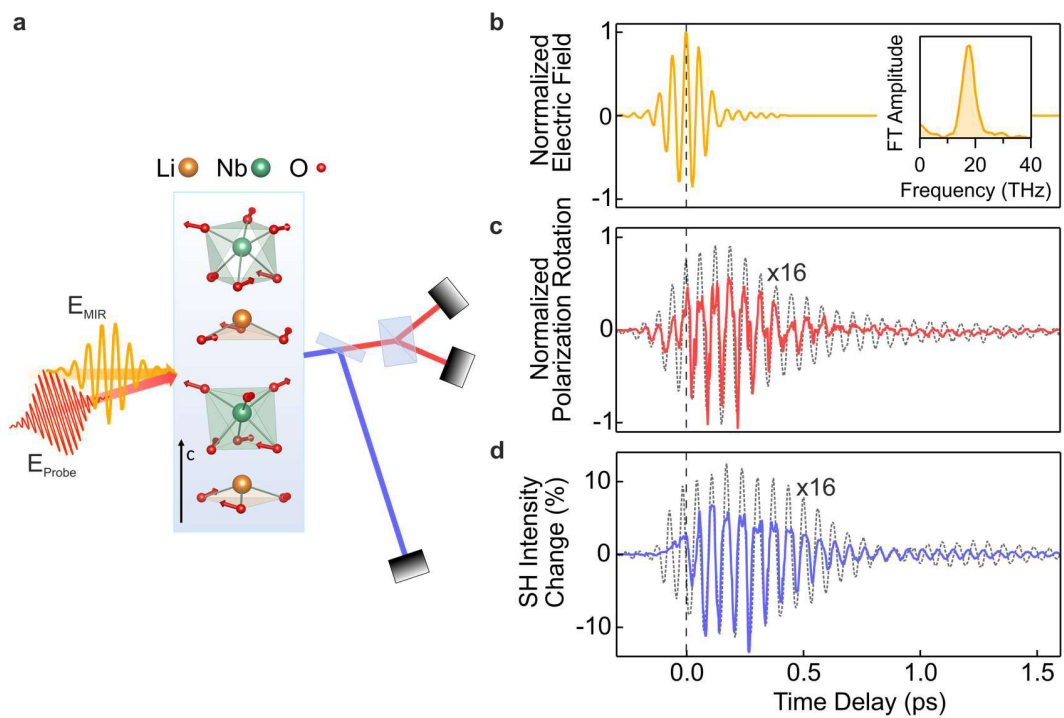

**Figure 1| Experimental setup and time-resolved optical changes. a**, schematic of the pump-probe configuration and the excited $A_1$ phonon mode in LiNbO$_3$, which has a net polar component along the crystal *c*-axis. **b**, electro-optic sampling measurement of the 150 fs, carrier envelope phase stable mid-infrared pulses, centered at 17.5 THz with 4 THz bandwidth, used for excitation of the $A_1$ mode. **c** and **d**, time resolved changes of the polarization rotation of the 800nm probe and second harmonic intensity, respectively, both for high (solid colored lines) and low (dashed lines) excitation fields.

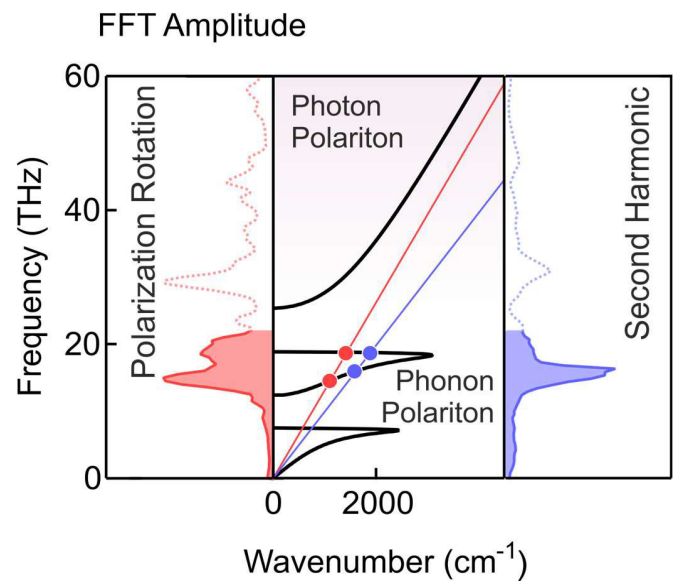

**Figure 2| Phonon-Polariton dispersion.** Phonon-Polariton dispersion in LiNbO$_3$ (black curve) and two lines representing the relation $\omega = v_g q$ for 800 nm (red) and 400 nm (blue) light. The dots mark the points of intersection with the dispersion relation, which precisely correspond to the observed fundamental frequencies (left and right panel).

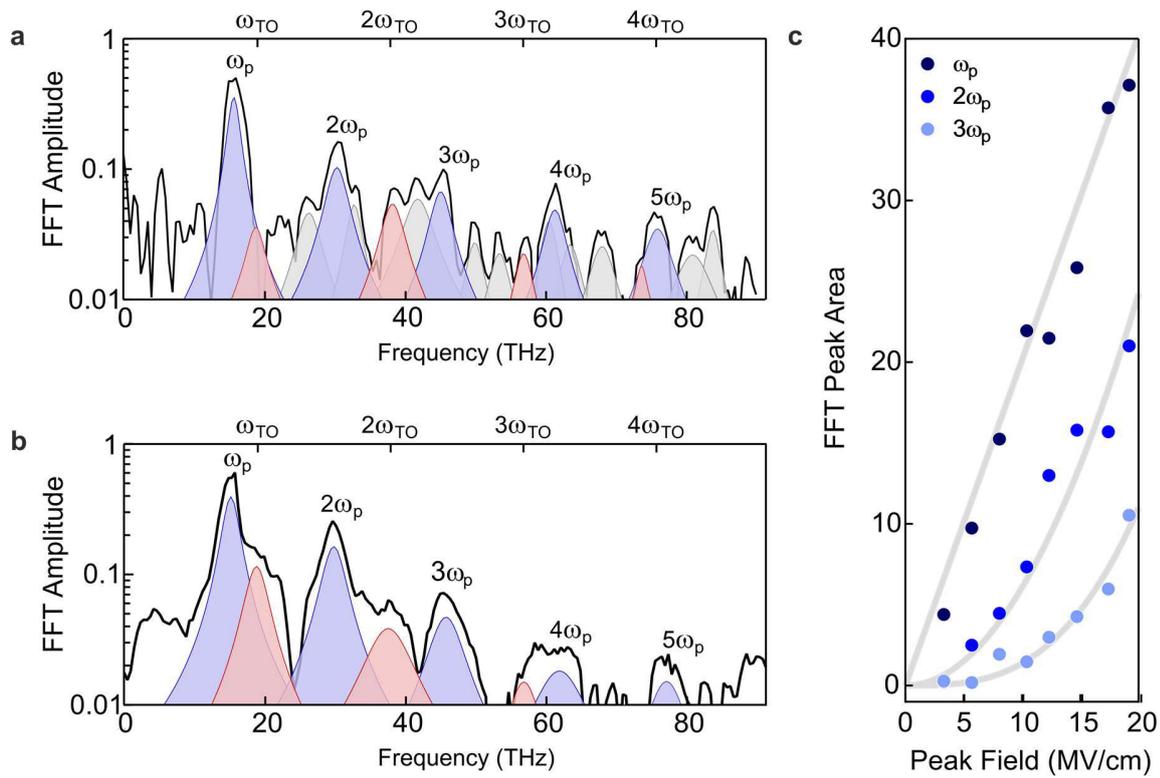

**Figure 3| Amplitude spectra and excitation field dependence. a** and **b**, FFT amplitude spectra of the polarization rotation and second harmonic intensity measurements, respectively, both for the high excitation field shown in Figure 1. The blue and read peaks correspond to multiples of the polariton frequency $\omega_p$ and $\omega_{TO}$. The fundamental phonon-polariton frequency for 800 nm and SH probe are 15.3 THz and 16.2 THz respectively. The grey peaks in panel **a** label sum and difference frequencies of $\omega_p$ and $\omega_{TO}$, which are absent in the SH response. **c**, Excitation field dependence of the peak area at the first, second and third harmonic of $\omega_p$, revealing a linear, quadratic and cubic trend.

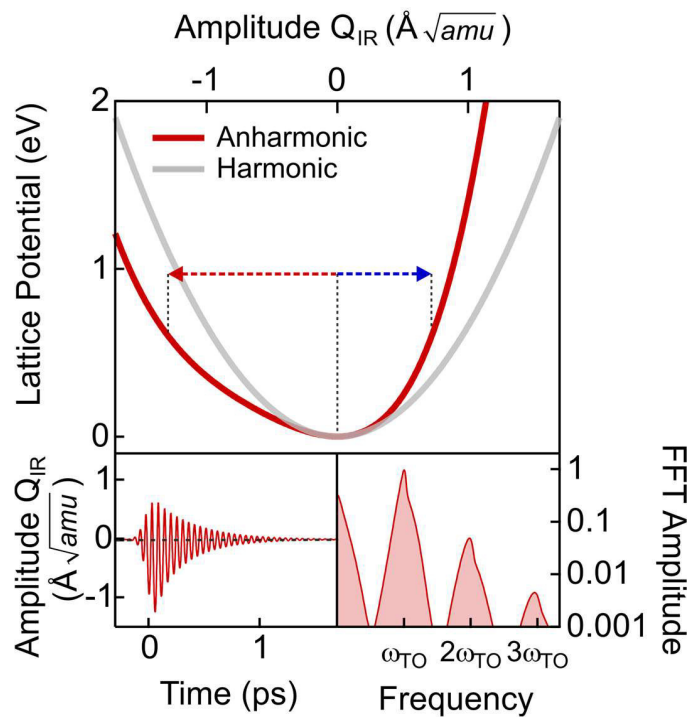

**Figure 4| Anharmonic lattice potential.** Lattice potential energy of LiNbO$_3$ (red) along the A$_1$ mode of Fig. 1a compared to a harmonic potential (grey) with the same fundamental frequency $\omega_{TO}$. The arrows denote the expected positive and negative excursions for an energy of 0.6 eV, roughly corresponding to the energy deposited per unit cell by the excitation pulses. The lower graph shows the solution of the equation of motion and its amplitude spectrum (see text).

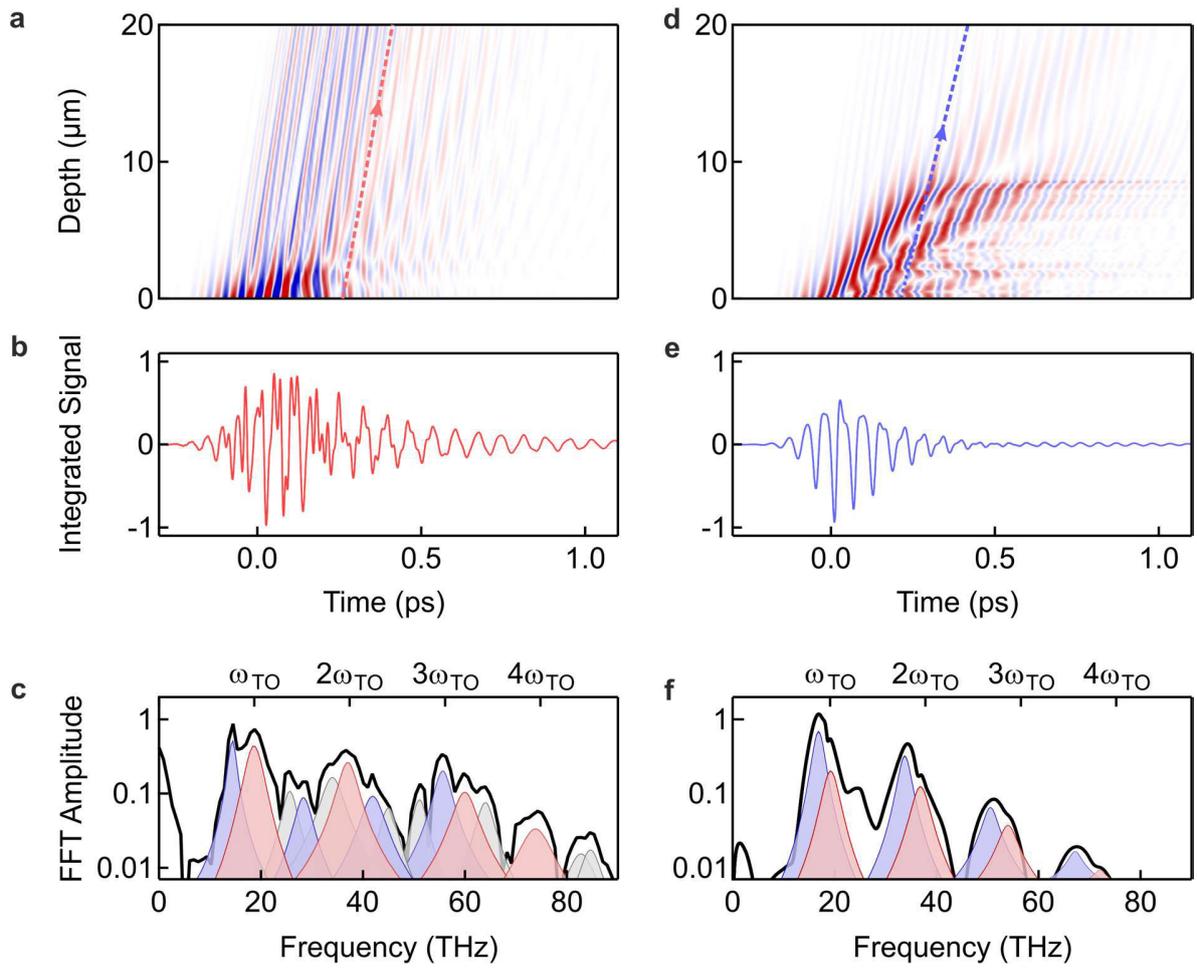

**Figure 5| FDTD phonon-polariton propagation simulations. a**, Electric field as a function of depth and time inside LiNbO$_3$ after MIR excitation, the red line shows the propagation of the 800 nm probe pulse for one time delay. **b,c** Time trace derived by integrating along the red line in **a** for all time delays and the corresponding amplitude spectrum. The spectrum shows harmonics of $\omega_p$ and $\omega_{TO}$ as well as mixed frequencies. **d**, Phonon-field $Q_{IR}$ from the same simulation as in **a**. The blue line shows the propagation of the SH light. **e**, Time trace derived from an integration along the blue line within the first 2 μm, where the SH light is generated. **f**, Amplitude spectrum of **e**, which only shows broad peaks at the harmonics of $\omega_p$ and $\omega_{TO}$, the mixed frequencies are absent.

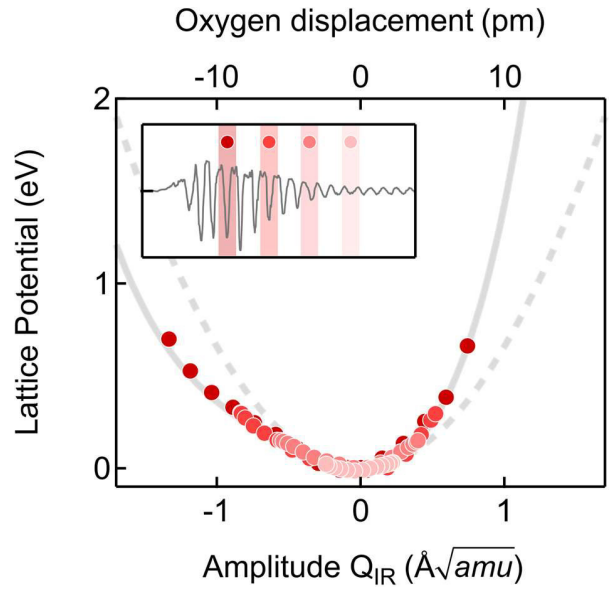

**Figure 6| Reconstructed $A_1$ mode potential energy.** Experimentally reconstructed potential energy of the $A_1$ mode (red colored circles) derived from different cycles of the experimental data, which are shown in the inset. The grey solid line is the mode potential obtained by DFT calculations. The experimental potential is scaled by a fit to the calculated potential using only one scaling factor with preserved aspect ratio. From this comparison, we estimate maximum mode excursions of 1.4 $\text{Å}\sqrt{amu}$, corresponding to a ~14 pm displacement of the oxygen atoms from their equilibrium positions. The dashed curve is the potential in the harmonic approximation.

**Acknowledgements:**

We thank Roberto Merlin for valuable discussions. The research leading to these results received funding from the European Research Council under the European Union's Seventh Framework Programme (FP7/2007–2013)/ERC Grant Agreement No. 319286 (QMAC). This work has been supported by the excellence cluster "The Hamburg Centre for Ultrafast Imaging—Structure, Dynamics and Control of Matter at the Atomic Scale" of the Deutsche Forschungsgemeinschaft.

Author Contributions

A.C. conceived this project together with A.v.H. and R.M.. R.M., A.v.H. and M. Först build the experimental setup. A.v.H. and R.M. conducted the experiment and analysed the data. M. Fechner performed the DFT calculations. All authors contributed to the manuscript.

# Methods

**Experimental Setup**

The carrier envelope phase stable 150 fs long 17.5-THz (4 Thz FWHM-bandwidth) mid-infrared pulses were obtained by difference frequency generation of two signal beams from two optical parametrical amplifiers, which were seeded by the same white light and pumped by 30-fs, 800-nm pulses at 1 kHz repetition rate. The dynamics were probed in non-collinear geometry with a 30° angle between the mid-IR pump and 800-nm probe pulses. (see Extended Data Fig. 1).

Due to the large second-order nonlinear susceptibility of $LiNbO_3$, the transmitted 800-nm pulses generated a second harmonic (SH) signal at 400 nm, which was separated from the fundamental beam after the sample using a dichroic mirror. Changes in the second harmonic intensity were measured with a photo multiplier tube. The probed SH dynamics originate from a layer of one coherence length $l_c$ = 1.3 µm below the surface (see supplementary information of Ref. 10 and Ref. 34 for details).

The pump induced polarization rotation of the 800-nm beam was detected by balancing the intensity on two photodiodes using a half-wave plate and a Wollaston prism. All experiments were conducted at room temperature and at ambient pressure. The sample used in the experiments was a commercially available $LiNbO_3$ single crystal (5x5x5 mm).

**Detection process of polarization rotation and second harmonic measurements**

The interaction of the probe beam with the phonon polariton can be descried as a nonlinear mixing of the oscillatory atomic motions with the probe beam, resulting in the generation of sidebands in the probe spectrum. These oscillations are detected by interfering the sidebands with a local oscillator on the detector[15,16]. The nonlinear mixing process is due to hyper-Raman scattering in the second harmonic measurements, and electro-optic mixing as well as Raman scattering in the polarization rotation measurements[13,14]. After interaction with a phonon-polariton of frequency $\Omega$, at a specific pump-probe time-delay $\tau$ the spectrum of the transmitted probe consists of two terms:

$$E(\omega) = E_0(\omega) + E_Q(\omega + \Omega) \cdot \exp(i\Omega \cdot \tau) + c.c.$$

The first term denotes the spectrum of the unperturbed probe beam and the second term accounts for the generated sideband with amplitude spectrum $E_Q(\omega + \Omega)$, which is proportional to the phonon amplitude. The sideband further acquires a time-delay dependent phase $\exp(i\Omega \cdot \tau)$, which is modulated at the phonon frequency. A phase sensitive measurement of the sideband thus carries information about both frequency and amplitude of the atomic motions involved in its generation (for a more detailed derivation see Ref. 31 and 32).

Both phase and amplitude can be detected by recording the time-delay dependent interference signal of the generated sidebands with transmitted unperturbed probe light. The recorded intensity on a detector is the spectral integral of this interference term:

$$I(\tau) = \int d\omega \left( \left| E_0(\omega) + E_Q(\omega + \Omega) \cdot \exp(i\Omega \cdot \tau) + c.c. \right|^2 \right)$$

$$= I_0 + A(\Omega)\cos(\Omega \cdot \tau) + B(\Omega)\cos(2\Omega \cdot \tau).$$

As the homodyne component $B(\Omega)$ is much smaller than the heterodyne component $A(\Omega)$, it can be disregarded.

Hence, the detected second harmonic light is sensitive to $I_0 + A(\Omega)\cos(\Omega\cdot\tau)$. The first term is a constant background, while the second term is proportional to $\Omega\cdot Q(\tau)$, where $Q(\tau)$ is the atomic motion (see Ref. 31 and 32). Generally, the proportionality constant depends on the phonon frequency due to different penetration depths. However, as the SH light was generated in a thin layer of 1.3 µm below the sample surface, its interaction length with the phonon-polariton harmonics was the same at all frequencies (see supplementary information of Ref. 10 and Ref. 34). The SH bandwidth supports efficient detection up to 60 THz (see Extended Data Fig. 3 a and b). A narrow bandpass filter on the high-energy side of the spectrum was used to shape the spectral response function and balance the detection efficiency of the first three harmonics[15] (15, 30 and 45 THz). We subtracted a slowly varying background from the experimental data to extract the oscillatory component. This background can be attributed to the modification of $\chi^{(2)}$ due to anharmonic phonon coupling to the ferroelectric soft mode (see Ref. 10). The resulting oscillatory changes in the SH measurements are directly proportional to the atomic motions of the excited high frequency $A_1$ mode of LiNbO$_3$ $Q(\tau)$.

The polarization rotation of the transmitted 800-nm light is measured by two balanced photodiodes placed behind a polarizing beamsplitter. Their difference signal is sensitive only to the heterodyne component $\Delta I(\tau) = I_{\parallel}(\tau) - I_{\perp}(\tau) = 2\cdot A(\Omega)\cos(\Omega\cdot\tau)$, which enhances the sensitivity to its oscillatory contribution. The large bandwidth of 90 THz of the 800 nm light was generated by pronounced spectral broadening due to self-phase modulation in the LiNbO3 crystal[35,36] as shown in Extended Data Fig. 3 c. The calculated sampling efficiency is shown in panel d. In contrast to the SH measurements, the interaction length between 800nm and phonon-polariton harmonics strongly depends on their frequency. Thus, these measurements cannot straight forwardly be used to derive the atomic motions. The grey curve in Extended Data Fig. 3 d displays the penetration depth, which significantly increases with ascending frequency.

**Assignment of all peaks in the PR amplitude spectrum**

Extended Data Figure 4 displays a detailed assignment of all peaks in the amplitude spectrum of the polarization rotation measurement. Circles/triangles represent a shift of the probe light by $\omega_p$ and $\omega_{TO}$, respectively. Multiple symbols indicate a shift by multiples of the fundamental frequency. Blue and red colors indicate up and down shifts, corresponding to sum and difference frequency mixing, respectively.

**Anharmonic frequency renormalization**

The black solid line in Extended Data Fig. 5 shows a copy of the amplitude spectrum of the second-harmonic measurement presented as Fig. 3b of the main text. The observed harmonics of the polariton frequency are slightly red-shifted with respect to integer multiples of the fundamental frequency (blue vertical lines). This effect is also observed in simulations of a resonantly driven anharmonic oscillator using the potential obtained from DFT calculations (red curve). This frequency renormalization is well known to occur for anharmonic oscillators at large oscillation amplitudes[28,29]. This behavior clearly distinguishes the discussed high-harmonic generation from cascaded Raman processes, which would produce equally spaced peaks in the spectrum.

**Linear Optical Properties**

The THz linear optical properties along the crystal c-axis of $LiNbO_3$ are dominated by two optical phonons at 7.8 THz and 18.9 THz. It further includes a weak mode at 8.2 THz and a feature at 21 THz which has been attributed to two phonon absorption[12]. The measured THz reflectivity spectrum of the investigated sample is shown in Extended Data Fig. 6 (grey line), together with a fit using four Lorentzian oscillators (red line). The resulting fit parameters for the two dominating optical phonons were used in the FDTD simulations of the phonon polariton propagation and are shown in Extended Data Table 1. Extended Data Figure 6 further shows the reflectivity fit considering only the two dominating oscillators only (blue line) and the reflectivity obtained from the FDTD simulation using these two modes (green line). Both yield a satisfactory description of the optical properties in the region of interest (12-20 THz).

**FDTD phonon polariton simulations**

The phonon-polariton propagation dynamics in LiNbO$_3$ have been calculated by solving Maxwell's equation in space and time. To this end we used a finite difference time domain approach in one spatial dimension[30].

To capture the linear response of the material we used the parameters obtained from Fourier transform infrared spectroscopy measurements and implemented two separate sets of equations to model the two dominant optical phonons along the crystal c-axis. For each mode, the equation of motion is given by

$$\ddot{Q}_{IR} + 2\gamma \dot{Q}_{IR} + \omega_{TO}^2 Q_{IR} = Z^* E(t).$$

Here, the damping term $\gamma$, the phonon frequency $\omega_{TO}$ and the born effective charge $Z^*$ are the experimentally obtained parameters. $Z^*$ can be expressed in terms of parameters derived from infrared spectroscopy as $\omega_{TO}\sqrt{(\varepsilon_0 - \varepsilon_\infty)}\sqrt{\frac{\epsilon_0}{n}}$, where $n$ is the oscillator density, $\varepsilon_0$ and $\varepsilon_\infty$ are the low and high frequency limit of the dielectric function respectively. Due to the presence of multiple modes, $\varepsilon_0$ and $\varepsilon_\infty$ had to be derived for every mode from the generalized Lydanne-Sachs-Teller relation[37]. The oscillator density was approximated as one oscillator per unit cell.

The above equation was solved at every discrete point of the grid in space and time using the calculated values of the electric field via Maxwell's equation. The oscillator equation and Maxwell's equation are coupled via the electric displacement field

$$D = \epsilon_0 \varepsilon_\infty E + \omega_{TO}\sqrt{(\varepsilon_0 - \varepsilon_\infty)}\sqrt{\epsilon_0\, n}\, Q_{IR}$$

By adding also the second oscillator equation to the dielectric displacement field, these relations reproduce the linear optical properties of LiNbO$_3$ (see Extended Data Fig. 6).

To describe the observed nonlinear effects, the lattice anharmonicities of the driven mode were introduced into the equation of motion of the excited A$_1$ mode at 19 THz:

$$\ddot{Q}_{IR} + 2\gamma \dot{Q}_{IR} + \omega_{TO}^2 Q_{IR} + a_3 Q_{IR}^2 + a_4 Q_{IR}^3 = Z^* E(t).$$

Here, $a_3$ and $a_4$ are the anharmonic constants taken from ab-initio Density Functional Theory calculations as described below ($a_3 = 1567.65\ meV/amu^{3/2}/\text{Å}^3$, $a_4 = 900.8\ meV/amu^2/\text{Å}^4$). Here, the mid-infrared pump pulse was set to a field strength of 30 MV/cm, carrier frequency 17.5 THz and 180 fs duration, comparable to the experiment. We evaluated the equations with a time stepping of 0.5 fs and spatial grid of 0.5 μm. At the boundaries of the simulated area, perfectly matched boundary conditions were implemented to impede back reflection.

**DFT calculations of the full anharmonic potential**

To explore the nonlinear response of a resonantly excited phonon mode we performed first-principle computations within the framework of density functional theory (DFT). All our computations were carried out using DFT as implemented in the Quantum Espresso code[38]. We used ultrasoft pseudopotentials, which contain as valence states the 2p 2s for Lithium, $4s^2 4p^6 4d^4 5s^1$ for Niobium and $2s^2 2p^4$ for Oxygen. As numerical parameters, we applied a cutoff energy for the plane wave expansion of 80 Rydberg and five times this value for the charge density. For all computations, we sampled the Brillouin zone by a 17x17x17 k-point mesh generated with the Monkhorst and Pack scheme[39] and reiterated total energy calculations until the total energy became less than $10^{-10}$ Rydberg. Before calculating phonon-modes we fully structural relaxed the unit-cell regarding forces and pressure below the threshold of 5 µRy/$a_0$. We finally performed density functional perturbation theory[40] calculations to obtain phonon modes eigenvectors and frequencies. Finally, we compute the anharmonic phonon potential by calculating the total energy for structures, which have been modulated with the phonon eigenvector. Least mean square fits of this total energy landscape reveal the anharmonic coefficients of eq. 2 of the main text and the phonon mode eigenvector as shown in Fig. 1a.

**Rescaling of the reconstructed potential**

In order to obtain the proportionality factor $B$ in $dI_{SH}/dQ = B \cdot \dot{Q}$, the experimentally reconstructed potentials are fitted by the potential derived from DFT calculations. With $dI_{SH}/dQ = B \cdot \dot{Q}$, $E_{kin} = \frac{1}{2}\frac{(dI_{SH}/dQ)^2}{B^2}$ and $Q = B \cdot \int (dI_{SH}/dQ)dt$. The constant $B$ can then be derived by fitting the function $f(Q) = B^2 \cdot U(Q/B)$ to the experimental data, where U(Q) is the potential obtained by DFT. With this we can rescale the experimental x and y axis and obtain the phonon amplitude in terms of $\text{Å}\sqrt{amu}$ and the potential energy in eV. The maximum displacement of the oxygen atoms involved in the vibration, can be calculated with the knowledge of the phonon eigenvectors which we obtained from DFT calculations. This calculation yields a maximum displacement of the oxygen atoms of approximately 14 pm which amounts to 7% of the Nb-O and 5% of the O-O nearest neighbor distance at the corresponding potential energy (0.7 eV), which agrees with the estimated energy deposited per unit cell (0.6 eV at 3 µJ pulse energy).

# Extended Data

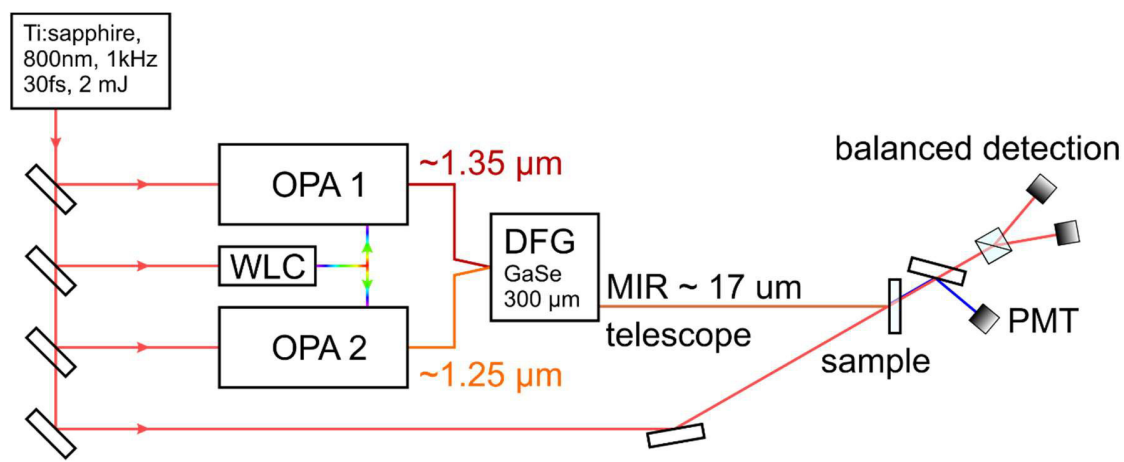

**Extended Data Figure 1** Schematic representation of the experimental setup. 30-fs pulses from a Ti:sapphire are used to pump two optical parametric amplifiers (OPA), which are seeded by the same white light continuum (WLC) . Carrier envelope phase stable 3-µJ 150-fs pulses at 17 µm wavelength are obtained by difference frequency generation of the two signal beams from the OPAs. The mid-infrared is focused to a spotsize of approximately 65 µm using a telescope and overlapped with the 800-nm probe beam (40 nJ, 35 µm spotsize).

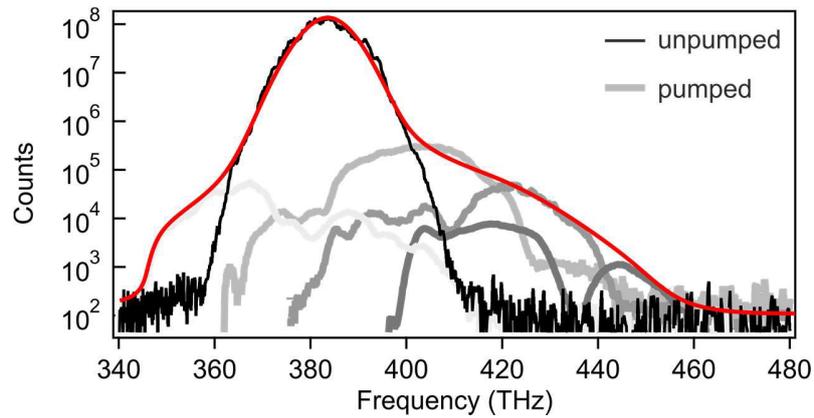

**Extended Data Figure 2** The black solid line is the incident spectrum of the 800-nm probe pulses with a bandwidth of ~ 30 THz. After interaction with the mid-infrared excited sample volume several sidebands are superimposed (grey lines). The red line is a guide to the eye of the resulting spectral broadening. Due to conservation of momentum the sidebands propagate in a slightly different directions compared to the unperturbed 800 nm. The differently shaded grey curves have been taken by scanning the half space behind the $LiNbO_3$ crystal.

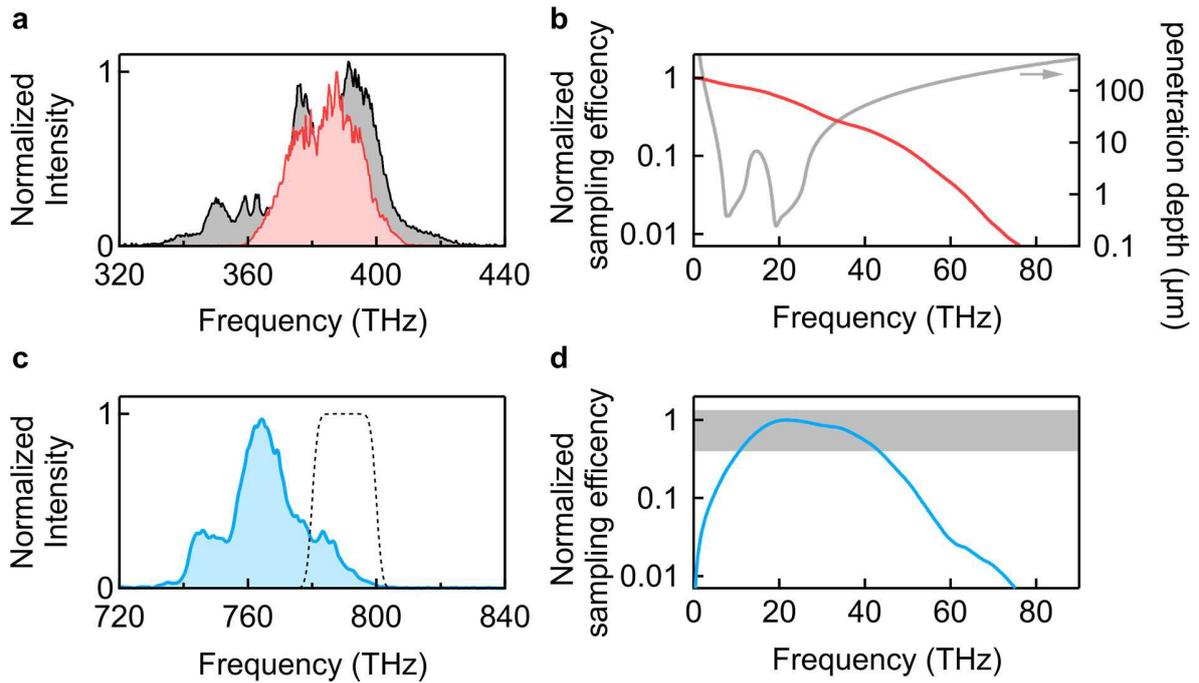

**Extended Data Figure 3 a**, Spectrum of the 800 nm probe beam before (red) and after (grey) propagation through the LiNbO3 crystal. **b**, sampling efficiency of the 800 nm light calculated with the spectrum in shown in **a** (red curve) and the penetration depth in the min-infrared region obtained from FT-IR spectroscopy. **c**, spectrum of the generated SH light and the acceptance range of the bandpass filter in front of the detector (dashed curve). **d**, sampling efficiency of the SH light with the spectrum shown in **c**. The sampling efficiency is almost constant in the region 15-45 THz allowing unperturbed measurement of the first three phonon harmonics.

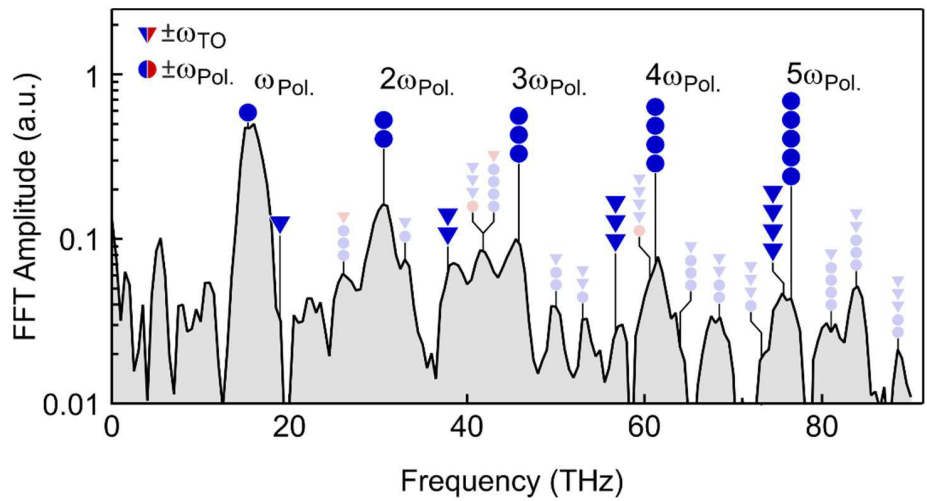

**Extended Data Figure 4** Amplitude spectrum of the polarization rotation measurement. Blue symbols denote a frequency blueshift of 15 THz (triangles) and 19 THz (circles). Multiple symbols represent a shift by multiples of the corresponding frequency. Red symbols denote a red shift.

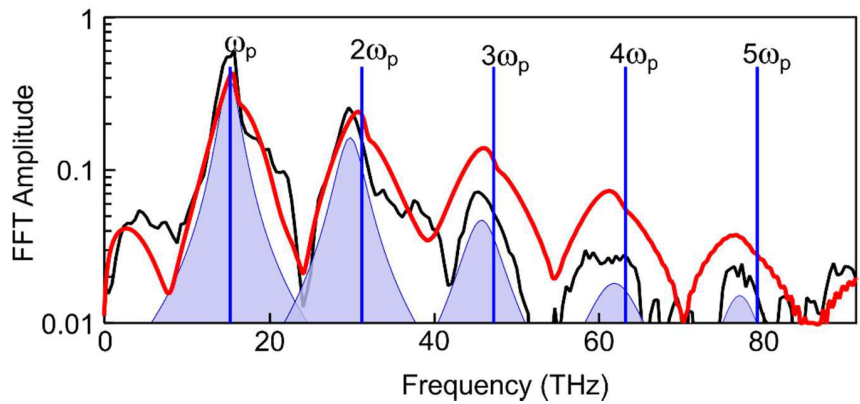

**Extended Data Figure 5** Amplitude spectrum of the second harmonic measurement (black curve). The blue lines denote multiples of the fundamental polariton frequency. The observed peaks associated with harmonics of the fundamental frequency are all shifted to the red with respect to the blue lines as expected for a strongly driven anharmonic oscillator. The red solid line is a simulation reproducing the red shift of the phonon harmonics.

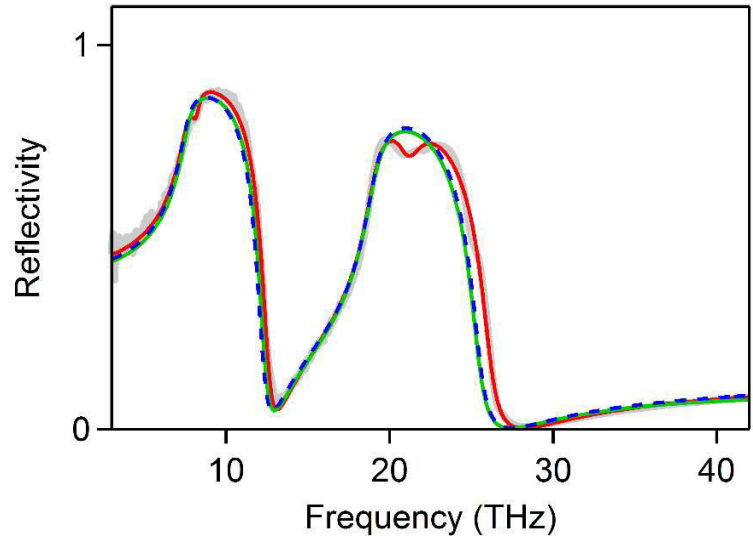

**Extended Data Figure 6** The grey solid line is the measured THz reflectivity spectrum of LiNbO$_3$ with light polarized along the c-axis. The red line is a fit to the data considering 4 Lorentzian oscillators. The dashed blue line is the fitted reflectivity only considering the two dominant phonon modes at 7.5 THz and 19 THz. The green line is the simulated reflectivity when only these oscillators are considered in the FDTD simulations (see Supplementary Information S6).

| Oscillator # | Frequency ($cm^{-1}$) | Oscillator strength($cm^{-1}$) $\omega_{TO}\sqrt{\varepsilon_0 - \varepsilon_\infty}$ | Damping ($cm^{-1}$) |
|---|---|---|---|
| 1 | 249.3 | 922.8 | 27.7 |
| 2 | 271.6 | 384.1 | 20 |
| 3 | 632 | 955.9 | 33.5 |
| 4 | 696.7 | 352.5 | 76.2 |
| $\varepsilon_\infty$ | 4.4054 | | |

**Extended Data Table 1** Values obtained from a fit of 4 Lorentzian oscillators to the reflectivity spectrum of our LiNbO$_3$ sample.